\documentclass[12pt, titlepage]{article}

\usepackage{xcolor}
\RequirePackage{amsthm,amsmath,amsfonts,amssymb,rotating,url}
\usepackage{pdfpages}
\usepackage{natbib}

\usepackage{amsmath,amssymb,amstext,amsthm,graphics, layout, lscape, longtable}
\usepackage[normalem]{ulem}
\usepackage{color}

\usepackage{xr}
\usepackage{url}
\usepackage{caption}
\usepackage{amssymb,amsmath,graphicx,multicol,epsfig,pictexwd,spverbatim}
\usepackage{color, colortbl}
\usepackage{multirow,multicol,dcolumn,booktabs}
\definecolor{mygray}{gray}{0.8}
\usepackage{lscape}
\usepackage[utf8]{inputenc}
\usepackage{graphicx}
\usepackage{hyperref}
\renewcommand{\baselinestretch}{1}

\usepackage{booktabs}
\usepackage{tabularx}
\usepackage{fourier}
\usepackage{comment}
\usepackage{tablefootnote}
\usepackage{tabu}
\usepackage{lscape}
\usepackage{array}
\usepackage{caption}
\usepackage{bm}
\usepackage{xcolor,colortbl}
\usepackage{multirow}
\usepackage{makecell}
\usepackage{longtable}
\usepackage{authblk}
\usepackage{tabularray}
\usepackage{subcaption}

\usepackage{array}
\usepackage{xcolor}


\newcommand{\bi}{\begin{itemize}}
\newcommand{\ei}{\end{itemize}}

\newcolumntype{C}{@{\extracolsep{.4em}}c@{\extracolsep{0pt}}}%

\usepackage{setspace}
\RequirePackage[top=1in, left=1in, right=1in]{geometry}
\begin{document}

\title{A Bayesian Learning Model for Joint Risk Prediction of  Alcohol and Cannabis Use Disorders}
\author[a]{Rajapaksha Mudalige Dhanushka S. Rajapaksha}
\author[a]{Tingfang Wang}
\author[a]{Thanthirige Lakshika M. Ruberu}
\author[b]{Joseph M. Boden} 
\author[a$^{*}$]{Pankaj K. Choudhary}
\author[a$^{*}$]{Swati Biswas}

\affil[a]{Department of Mathematical Sciences, University of Texas at Dallas, Richardson, TX 75080, United States}
\affil[b]{Department of Psychological Medicine, University of Otago, Christchurch 8011, New Zealand}
\affil[*]{Correspondence to Swati Biswas, 800 W Campbell Rd, FO 35, Richardson, TX 75025, swati.biswas@utdallas.edu OR Pankaj K. Choudhary, 800 W Campbell Rd, FO 35, Richardson, TX 75025, pankaj@utdallas.edu}
\date{}
\maketitle

\begin{abstract} Substance use disorders (SUDs) are a serious public health concern in the United States. Alcohol and cannabis are two of the most widely used substances. For adolescent/youth users of alcohol or cannabis, we propose a joint Bayesian learning model to predict their risks of developing alcohol use disorder (AUD) and cannabis use disorder (CUD) in adulthood based on their personal risk factors. The model is trained on nationally representative longitudinal data from Add Health ($n = 12,503$). It consists of sub-models that predict the two SUDs for three groups of users --- those who use alcohol only, cannabis only, and both substances --- based on shared as well as unique risk factors. The model comprises of ten predictors. We externally validate the model on two independent datasets. The areas under the receiver operating characteristic curves for AUD and CUD, respectively, are: (a) 0.719 and 0.690 based on 5-fold cross-validation, (b) 0.748 and 0.710 based on validation dataset 1, and (c) 0.650 and 0.750 based on validation dataset 2. A simulation study shows that the proposed joint modeling approach generally performs better than separate univariate modeling of the corresponding dependent outcomes in terms of predictive accuracy. Our model may help in identifying adolescent substance users at high risk of developing SUD in adulthood, who can then be helped with appropriate intervention. 
\vspace{1em} 

\end{abstract}

\section{Introduction}

Substance use disorders (SUDs) are an ongoing public health crisis in the United States (US). In 2021, among people aged 12 or older, 46.3 million (16.5\%) had a past year SUD, including 29.5 million (10.6\%) with alcohol use disorder (AUD) and 16.3 million (5.8\%) with cannabis use disorder (CUD) (SAMHSA,~2022). \nocite{SAMHSA2022}  Besides causing  human suffering, substance abuse and SUDs also have serious economic consequences, costing more than \$740 billion annually in crime, lost work productivity, and health care (NIDA,~2023). \nocite{NIDA2023} Most adults who develop SUDs start substance use in adolescence/youth (CDC,~2023). \nocite{CDC2023} In 2018, the lifetime alcohol and cannabis use prevalence among students in grades 8-12 were 41.2\% and 29.7\%, respectively \citep{Johnston2019}.  It is also common for a person to use two or more substances and for a person with an SUD to have two or more disorders \citep{Yurasek2017}.
%It is also common for a person to use two or more substances (CDC,~2022) \nocite{CDC2022} and for a person with an SUD to have two or more disorders \citep{Yurasek2017}.
%Polysubstance use and disorders are also common (CDC,~2022). \nocite{CDC2022} It is quite common for a person with an SUD to have two or more disorders \citep{Yurasek2017}. 
Out of the 46.3~million people mentioned above, 7.3 million (2.6\%) had SUDs of both alcohol and drugs (SAMHSA,~2022). \nocite{SAMHSA2022} Among adolescents, 34\% had used two or more of alcohol, cigarettes, and cannabis prior to age 16 \citep{Moss2014}. 

These alarming statistics, together with the fact that substance use in adolescence is a stepping stone to SUDs in adulthood, underscore the importance of identifying high risk adolescent users for providing early intervention. %\citep{Afuseh2020}. 
For this task, risk prediction models are needed that can take the individual-level risk factors as input and provide personalized, future quantitative risk of developing SUD in adulthood as output. Such models are in clinical use for various diseases, including breast cancer and heart disease \citep{DAgostino2008, bcrat, Chowdhury2017, Cattelani2019, Caye2019}.
Some tools have been proposed for SUDs as well \citep{Hayatbakhsh2009, Meier2016, Jing2020, %Hu2020, %Rajapaksha2020, Nasir2021
Rajapaksha2022, ZhangJames2020}. 
However, with the exception of the \cite{Rajapaksha2022} model, none have been built using a large, nationally representative longitudinal dataset containing a comprehensive set of SUD risk factors with follow-up from adolescence to adulthood and none have been externally validated. Thus, their practical utility is limited. The Rajapaksha et al.\ model predicts for a cannabis user the risk of developing CUD in adulthood based on their personal risk factors measured in adolescence/youth. The model was trained on data from the National Longitudinal Study on Adolescent Health (Add Health) \citep{Harris2009} using a  Bayesian learning approach. Its external validation on three independent test datasets established not just the good performance of the model, but also its robustness and portability \citep{Rajapaksha2022, Ruberu2023}.

Although there has been some interest in developing risk prediction models for a single SUD, little attention has been paid to joint models for multiple SUDs. One exception is \cite{Ruberu2022}, in which the authors proposed a joint model for predicting risks of hazardous use of alcohol, cannabis, and tobacco for adolescent users of all three substances. But it is a preliminary model because the data used for training were cross-sectional and not nationally representative. Moreover, its target population is users of all three substances, which restricts the applicability of the model to a relatively small group. To our knowledge, no other joint risk prediction model has been developed for two or more SUDs. 

Of specific interest in this article is joint modeling of AUD and CUD --- two widely prevalent SUDs. They have some shared etiology and risk factors \citep{Crane2021, Hatoum2023}
%, which makes them dependent. 
Moreover, from a clinical perspective, it is important to identify substance users at high risk of either disorder. In addition, the risk factors for the two SUDs and their effects are likely to differ across three groups of users: those who use only alcohol (A), both alcohol and cannabis (B),  and only cannabis (C).  In this context, joint modeling offers some advantages over separate substance-specific univariate modeling: (a) borrowing of information across the three groups and the two outcomes, potentially leading to higher predictive accuracy; (b) providing larger sample size and wider applicability of the model by targeting users of at least one of the two substances rather than users of only one substance or both substances; and (c) leading to deeper insights into the common mechanisms underlying the two disorders. 

For adopting a risk prediction model in practice, especially one that targets a sensitive population such as the adolescent substance users, model parsimony is essential. To this end, regularization can help by shrinking regression coefficients towards zero. It also reduces overfitting and improves predictive accuracy of the model on future unseen data. For joint modeling with regularization, Bayesian learning offers an attractive framework because the model can be specified via conditionally independent components and regularization can be applied through appropriate prior choice \citep{VanErp2019}.  Moreover, unlike its classical counterparts, a Bayesian approach allows the regularization parameters to be estimated simultaneously along with the model parameters, thereby automatically accounting for their estimation uncertainty through posterior distributions \citep{Park2008, VanErp2019, OHara2009}.

In this article, we build a joint model for predicting risks of AUD and CUD in adulthood using risk factors from adolescence and validate the model. We proceed in three steps. First, we develop a Bayesian learning methodology to jointly model AUD and CUD for users of at least one of the two substances. The model has novel features necessitated by the specifics of the data and the problem at hand. For example, among the three user groups mentioned before, group A (alcohol-only users) is at risk for AUD but not CUD; group B (users of both alcohol and cannabis) is at risk for both AUD and CUD; and group C (cannabis-only users) is at risk for CUD but not AUD. To let the risk depend on the outcome-group combination, the joint model has $2 \times 3 = 6$ sub-models, allowing the predictors and their effects to vary with the sub-model. In particular, the sub-models for CUD in group~A and AUD in group~C should not have any predictors because the probability of developing an SUD is zero for a non-user of the corresponding substance. In the second step, we apply the proposed methodology to train a model on the Add Health data. Lastly, we externally validate the trained model on two independent test datasets. 
 
The remainder of this article is organized as follows. In Section~\ref{se:Add Health}, we introduce the Add Health data. Section~\ref{se:methods} describes the proposed joint modeling methodology. The methodology is then applied to the Add Health data to build a risk prediction model for AUD and CUD in Section~\ref{se:results}. In Section~\ref{se:validation}, we validate the model on two external datasets: one is an independent test set from Add Health and the other is from the Christchurch Health and Development Study (CHDS) \citep{Poulton2020}. 
In Section~\ref{se:simulation}, we evaluate properties of the joint modeling approach and compare with a standard univariate modeling approach via a simulation study. Section~\ref{se:discussion} concludes with a discussion. A web supplement is provided for additional details.

\section{Add Health Training Data} \label{se:Add Health}

% Move this para to supplement %%
%%Add Health conducted its first survey (wave~I) during 1994--95 school year when the participants were in grades 7--12 \citep{Harris2009}. Participants were selected using a multistage stratified cluster sampling design to ensure that they were representative of the then US adolescent population in terms of urbanicity, region, school size and type, and ethnicity. Follow-up surveys were conducted in 1996 (wave~II), 2001--02 (wave~III), 2008 (wave~IV), and 2016--18 (wave~V). No substance use related information was collected in the last wave. So the relevant data for our purpose come only from the waves~I--IV during which the participants became adults aged 24--32. The outcome variables --- binary measures of lifetime AUD and CUD diagnosed using the 4th edition of the Diagnostic and Statistical Manual of Mental Disorders \citep{Marel2019} --- were measured only in wave~IV. 
%%%%%

%We processed the data along the lines of \cite{Rajapaksha2022} to prepare a training set (see Supplement Section~\ref{se:training data} for details). 
The data from the Add Health study \citep{Harris2009} were processed along the lines of \cite{Rajapaksha2022} to prepare a training set (see Supplement Section S1 for details). The set consists of complete data from $n=12,503$ participants on 21 potential predictors (15 for AUD, 17 for CUD, and 11 shared by both). The predictors are listed in Supplement Table S1. %Of these, 15 are for AUD while 17 are for CUD (11 among these are shared). 
They include demographic factors such as gender and race, personality scores, physical and mental health indicators, measures of early life stress, and peer and familial factors such as peer substance use and parental education. For participants who developed AUD or CUD, only the information up to their age of onset was used (those who developed both SUDs had  their age of onset defined as the minimum of the two ages). This ensures that the predictor values precede the response and are not effects of SUDs.

Table~\ref{ta:new1} shows the composition of the training set and the lifetime prevalences of AUD and CUD. Most of the participants (64\%) are in group~B, followed by group~A (33\%), and group~C (3\%). The prevalence of SUD varies widely with the disorder and the group. Overall, the prevalence of AUD (9.5\%) is higher than that of CUD (5.1\%). As expected, AUD prevalence in group~C and CUD prevalence in group~A are zero. Both prevalences are highest in group~B: 12.5\% for AUD and 7.8\% for CUD. Moreover, AUD prevalence in group~A and CUD prevalence in group~C are about 35\% less than their respective prevalence in group~B. The sample correlation between AUD and CUD is 0.19 and the chi-square test of association gives practically zero $p$-value based on all the data as well as the data restricted to group~B. Thus, there is a clear evidence of dependence between the two outcomes. 

% Table 1
%%%%%%%%
\begin{table}[ht]
  \caption{Composition of Add Health training data and prevalences of AUD and CUD.}
  \label{ta:new1}
  \begin{center}
  \begin{tabular}{lrrr} \hline
   \textbf{Group} & \textbf{Size} & \textbf{AUD}  & \textbf{CUD}  \\ \hline
A (alcohol-only users) & 4107 & 4.48\% & 0  \\
B (users of both) & 8013 & 12.48\% & 7.80\% \\
C (cannabis-only users) & 383 &  0  & 2.87\% \\ \hline
Overall & 12503 & 9.47\% & 5.08\%  \\ \hline
 \end{tabular}
\end{center}
\end{table}
%%%%%%%%%%

All quantitative predictors are scaled within [0,1] to allow portability across datasets \citep{Rajapaksha2022}. Supplement Table S2 presents the characteristics of SUD cases and controls (i.e., the non-SUD participants) with respect to the cross-sectional predictors. Compared to controls, the AUD cases are more likely to be males (60\% vs 46\%); experienced more adverse childhood experiences (ACEs) on average (0.23 vs 0.20); and reported higher neuroticism (0.55 vs 0.52) and openness (0.75 vs 0.73) but lower conscientiousness (0.70 vs 0.73). The cases are also more likely to be from white race (82\% vs 68\%); reported more extraversion on average (0.70 vs 0.66); and had more educated parents (0.69 vs 0.64) but were more likely to stay away from school troubles (0.31 vs 0.26). Likewise, the CUD cases are more likely to be males (61\% vs 47\%); experienced more ACEs on average (0.26 vs 0.20); and reported higher neuroticism (0.56 vs 0.52) and openness (0.76 vs 0.73) but lower conscientiousness (0.70 vs 0.73). 

Supplement Tables S3  and S4 present sample characteristics of the longitudinal predictors selected for AUD and CUD outcomes, respectively. In general, average peer alcohol use, peer smoking, peer cannabis use, and depressive symptoms increase over time, while delinquent activities, self esteem, and violence victimization decrease. For both AUD and CUD, the average value of a predictor is at least as high for cases as for controls for all predictors, except for supportive environment.

\section{Joint Bayesian Modeling of AUD and CUD} \label{se:methods}

Let the index $j = \text{A, B, C}$ denote the three substance user groups. Of the $n$ participants in the data, suppose $n_j$ belong to group~$j$. They are indexed as $i=1, \ldots, n_j$ and $n_{\text{A}}+n_{\text{B}}+n_{\text{C}} = n$. Also, let $k$ index the SUDs: $k=1$ for AUD and $k=2$ for CUD. Further, let $y_{ij}^k$ be the indicator of SUD~$k$ for participant~$i$ in group~$j$ serving as the response. Since group~A is not at risk of CUD and group~C is not a risk of AUD, we have $y_{iA}^2 = 0 = y_{iC}^1$ for all participants.

Next, let $\textbf{x}_{ij}^k$, ${\beta}_{0j}^k$, and $\pmb{\beta}_j^k$ denote the predictor vector of participant~$i$ in group~$j$ for SUD~$k$ and the associated regression intercept and slope vector, respectively. The predictor vectors include both cross-sectional predictors and participant-specific random intercepts and slopes based on a linear mixed effects model (LMM) fit representing the longitudinal predictors (see Supplement Section S1). Note that the predictor vectors  and their regression coefficients are allowed to differ not only by the outcome $k$ but also by the group $j$. Although there may be an overlap between predictors for two or more groups/outcomes, the shared predictors can have different effects (regression coefficients). 

\subsection{Likelihood Model}

Add Health used a complex sampling design with region (of the US) as a stratification variable, which we include as a covariate in $\textbf{x}_{ij}^k$. There is also a clustering variable, school, as the participants are nested within schools. To account for this nesting, we include a school-level random intercept in the model that is shared by all participants within a school.\footnote{The school-level random intercept was included even in the LMMs for longitudinal predictors.} Let $u_{s[i,j]}$ be the random effect of the school to which participant~$i$ in group~$j$ belongs. Next, to account for dependence between the two SUD outcomes for the same participant, we also introduce a participant-level random intercept $u_{ij}$. Thus, the data are modeled as follows. For $i=1, \ldots, n_j$, $j=\text{A, B, C}$,  and $k=1,2$, 
\begin{equation}  \label{eq:conditional model}
y^k_{ij} | \textbf{x}_{ij}^k, \beta_{0j}^{k}, \pmb{\beta}_j^{k}, u_{ij}, u_{s[i,j]}  \sim \text{independent Bernoulli}(p^k_{ij}), 
\end{equation}
where $p^k_{ij}$ is the conditional probability of SUD~$k$ for participant~$i$ in  group~$j$. Its logit is assumed to have the form:
\begin{equation}
\text{logit}(p^k_{ij}) =  \beta_{0j}^{k}+(\textbf{x}_{ij}^{k})^T \pmb{\beta}_j^{k} + u_{ij} + u_{s[i,j]},  \label{eq:prob}
\end{equation}
where 
\begin{equation}
u_{ij} | \sigma_u \sim \text{independent } N(0,\sigma^2_u) \text{ and } u_{s[i,j]} | \sigma_{s} \sim \text{independent } N(0,\sigma^2_{{s}}),  \label{eq:random effects}
\end{equation}
with $u_{ij}$ and $u_{s[i, j]}$ being mutually independent. This model has six sub-models --- one for each outcome-group combination --- and each has its own set of predictors and regression coefficients. As noted previously, group~A is not at risk of CUD and group~C is not at risk of AUD, or equivalently, the predictors for these two group-outcome combinations have zero effects. To accommodate this particularity, we assign priors for their slope coefficients that are concentrated near zero (described later).

Add Health also provides survey weights to account for its complex sampling design. Let $w_{ij}$ denote the weight for participant~$i$ in group~$j$. The weights are scaled so that $\sum_i \sum_j w_{ij} = n$. Also, let $\textbf{u}$, $\textbf{u}_{s}$, $\pmb{\beta}_0$, and $ \pmb{\beta}_1$ denote the vectors of all participant random effects, school random effects, intercept terms, and slope coefficients, respectively. Further, let $\textbf{x}$ and $\textbf{y}$ denote the vectors of all predictor and response values. The weighted conditional log-likelihood of the data \citep{savi:toth:2016} under model~\eqref{eq:conditional model} is given as
\begin{equation} \label{eq:log-likelihood}
\log\{f(\textbf{y} | \textbf{x}, \pmb{\beta}_0, \pmb{\beta}_1, \textbf{u}, \textbf{u}_{s})\} = \sum_{j = A}^{C} \sum_{i = 1}^{n_j} \sum_{k = 1}^2  w_{ij} \big\{ y_{ij}^k \log \big(p_{ij}^k \big) + \big(1 - y_{ij}^k \big) \log \big(1 - p_{ij}^k \big) \big\}, 
\end{equation}
where $p_{ij}^k$ is given by~\eqref{eq:prob} and $f(\pmb{a} | \pmb{b})$ is the conditional probability function of $\pmb{a} | \pmb{b}$.

\subsection{Prior Distributions} \label{se:priors} 

We assign the following priors for the model parameters: 
\begin{gather*}
{\beta}_{0j}^{k} \sim  N(0,100), ~ \pmb{\beta}_j^k \sim \text{shrinkage priors}, ~ j = \text{A, B, C}; \, k = 1, 2, \\
    \sigma_u \sim \text{ Half-Cauchy} (0,4), ~ \sigma_{{s}} \sim \text{ Half-Cauchy} (0,10),
\end{gather*}
and assume the distributions to be mutually independent. The  variance of 100 for the intercept terms and the scale parameter of 10 for $\sigma_s$ make the respective priors non-informative. The scale parameter of 4 for $\sigma_u$ leads to approximately 0.90 prior probability for $\sigma_u$ between 0 and 25, a realistic range based on our data. Two shrinkage priors, lasso and $t$, are considered for the slope vectors. In particular, let $\pmb{\lambda}_j^k$ be the vector of shrinkage parameters associated with $\pmb{\beta}_j^k$ element-wise. Further, let $\pmb{\beta}_{lj}^k$ and $\pmb{\lambda}_{lj}^k$ denote a general element of the two vectors. Then, the two shrinkage priors are respectively given as
\[ 
\beta^k_{lj} | \lambda^k_{lj} \sim \text{Double-Exponential} \big(0, 1/ \lambda^k_{lj}\big), ~~ \beta^k_{lj} | \lambda^k_{lj} \sim  \text{Student's-$t$} \big(1, 0 , 1/\lambda^k_{lj} \big),
\]
where in both cases
\[
\lambda^k_{lj} \sim 
\begin{cases}
\text{Half-Cauchy} (0, 5), & (j, k) = (\text{A}, 2) \text{ or } (\text{C}, 1),  \\ 
\text{Half-Cauchy} (0, 1), & \text{otherwise}. 
\end{cases}
\]
The different elements of both $\pmb{\beta}_j^k$ and $\pmb{\lambda}_j^k$ are assumed to be mutually independent. The scale parameter of 5 for $\lambda^k_{lj}$ makes the marginal prior of $\beta^k_{lj}$ concentrated near zero. This reflects the {\em a priori} knowledge that the probability of getting SUD for a non-user of the substance, specifically, CUD in group~A and AUD in group~C, is zero. The scale parameter of 1 for the remaining group-outcome combinations is a standard choice \citep{VanErp2019}.

\subsection{Posterior Distributions} 

The unknowns in the model are the regression coefficient vectors $\pmb{\beta}_0$ and $\pmb{\beta}_1$, random effect vectors $\textbf{u}$ and $\textbf{u}_{s}$, standard deviations (SDs) $\sigma_u$ and $\sigma_{s}$, and the vector $\pmb{\lambda}$ of all shrinkage parameters. For their joint posterior, we can write,
\begin{align*}
f(\pmb{\beta}_0, \pmb{\beta}_1, \textbf{u}, \textbf{u}_{s},  \sigma_u, \sigma_{s}, \pmb{\lambda} | \textbf{y}, \textbf{x}) & \propto  f(\textbf{y} | \textbf{x}, \pmb{\beta}_0, \pmb{\beta}_1, \textbf{u}, \textbf{u}_{s}) f(\pmb{\beta}_0) f(\pmb{\beta}_1 | \pmb{\lambda}) f(\textbf{u} |\sigma_u) \\ 
& \hspace{3em}  \times  f(\textbf{u}_{s}|\sigma_{s})   f(\sigma_u) f(\sigma_{s}) f(\pmb{\lambda}),
\end{align*}
where the first term on the right is the exponential of the log-likelihood in~\eqref{eq:log-likelihood}. Since this posterior does not have a closed-form, Bayesian inference is performed using posterior simulation via a Markov Chain Monte Carlo algorithm. We use {\tt RStan} package for this task \citep{Stan2022}.

\subsection{Predicted Probability for a New Individual} \label{se:prediction}

Let $\tilde{p}_j^k$ be the marginal probability of SUD~$k$ under model~\eqref{eq:conditional model} for an individual who belongs to substance use group~$j$ and has predictor $\textbf{x}^k_j$. This probability is a function of model parameters $(\beta_{0j}, \pmb{\beta}_j^k, \sigma_u, \sigma_{s})$ obtained by applying the inverse logit transformation on~\eqref{eq:prob} and averaging over the two random effects. For given parameter values, the probability can be calculated numerically using Gauss-Hermite quadrature \citep{Lange2010}. 

Now, consider a new (future) individual belonging to group~$j$ whose risk prediction is needed. Let $\textbf{x}^k_{0j}$ and $y^k_{0j}$ be their predictor vector and response value for SUD~$k$. The predicted probability of SUD~$k$ for this individual is the posterior predictive probability $p^k_{0j} = P(y^k_{0j} = 1| \textbf{y}, \textbf{x}^k_{0j})$, which equals the posterior mean $E(\tilde{p}_j^k | \textbf{y}, \textbf{x}^k_{0j})$ of $\tilde{p}_j^k$ and can be computed using the posterior draws of $(\beta_{0j}, \pmb{\beta}_j^k, \sigma_u, \sigma_{s})$.

\subsection{Variable Selection and Model Comparison} \label{se:variable selection}

The shrinkage priors used for regression slopes are continuous and hence do not automatically lead to variable selection. Therefore, we apply two common Bayesian variable selection methods \citep{VanErp2019}: credible interval and thresholding (see Supplement Section S2). The supplement also describes how the models are compared on the basis of three predictive accuracy measures estimated using 5-fold cross-validation (CV), separately for AUD and CUD: area under the receiver operating characteristic curve (AUC); the ratio E/O of expected number (E) and observed number (O) of cases; and Brier score.

\section{Training of the Joint Model for CUD and AUD}\label{se:results}

First, we fit the joint model to the Add Health training data with 21 predictors --- 15 for AUD and 17 for CUD --- using lasso and $t$ shrinkage priors (Section~\ref{se:priors}). Then, we apply the credible interval and thresholding methods to perform variable selection for each sub-model based on predictive accuracy measures for AUD and CUD calculated via 5-fold CV (Section~\ref{se:variable selection}). The marginal probability $\tilde{p}^k_j$ needed for a prediction is calculated using Gauss-Hermite quadrature (Section~\ref{se:prediction}).  We tried bivariate quadrature as well as univariate quadrature by omitting the school effect $u_s$ because its variance estimate is close to zero (posterior mean = 0.145, 95\% credible interval: (0.031, 0.290)). As the results are practically the same, we use the latter, a simpler method. 

The thresholding method with 0.10 cutoff gives a joint model with highest predictive accuracy with both shrinkage priors. The accuracy estimates are reported in Table~\ref{t:ch4t5} as ``Original.'' We further explore some parsimonious and simpler versions of these models. Their accuracy estimates are also reported in the table. First, to enhance the clinical utility of the models, we exclude the stratification variable region and find that it does not affect predictive accuracy. Then, we explore the possibility of including the longitudinal predictors in the model simply as averages over the multiple waves instead of LMM-based random effects. We find that the former, in fact, yields higher accuracy. Moreover, although both priors lead to comparable model performance, those fitted using $t$ priors have a slightly higher AUC for predicting AUD compared to their lasso counterparts. Lastly, in the model with $t$ prior, even after excluding the predictor ACE (because its 95\% credible interval contains zero) and using a dichotomous version of race (white vs non-white), we get essentially the same accuracy.  Therefore, we use this simpler model as our final proposed model. For predicting AUD and CUD, its respective AUCs are 0.719 and 0.690; E/O are 0.994 and 0.987; and Brier scores are 0.080 and 0.046. On the whole, the model shows good calibration and discrimination performance. 

% Table 2

%%%%%%%%%%%%%%%%%
\begin{table}
  \caption{Comparison of prediction performance of various joint models obtained using 10\% threshold. Average AUC and E/O are obtained as average over the five folds.}
   \label{t:ch4t5}
  \begin{center}
  \begin{tabular}{cp{1.75in}cccccccc} \hline
    & &   \multicolumn{2}{c}{\textbf{AUC}} &&  \multicolumn{2}{c}{\textbf{E/O}} && \multicolumn{2}{c}{\textbf{Brier score}} \\ \cline{3-4} \cline{6-7} \cline{9-10}
 \textbf{Prior} & \textbf{Model} & \textbf{AUD} & \textbf{CUD} && \textbf{AUD}  & \textbf{CUD} && \textbf{AUD}  & \textbf{CUD} \\ \hline
  lasso &  Original & 0.702  & 0.671 && 1.012  & 0.977 && 0.081 & 0.046 \\
  & Region excluded & 0.703  & 0.671 && 0.998  & 0.978 && 0.081 & 0.046 \\
  & Average for longitudinal predictors & 0.713  & 0.690 && 1.001  & 0.988 && 0.081 & 0.046 \\[0.4em] \hline
 $t$ & Original&  0.704  &0.671 && 1.013  & 0.978 && 0.081 & 0.046\\
  & Region excluded & 0.705  & 0.671 && 0.998  & 0.977 && 0.081 & 0.046 \\
  & Average for longitudinal predictors & 0.718  & 0.689 && 1.001  & 0.987 && 0.081 & 0.046\\
  & ACE$^*$ excluded and race dichotomized &  0.719  & 0.690 && 0.994  & 0.987 && 0.080 & 0.046\\ \hline
   \end{tabular}
  \end{center}
$^*$ adverse childhood experience  
\end{table}
%%%%%%%%%%%%%%%%%

Table~\ref{t:ch4t6} presents the predictors in the sub-models of the final model. As expected, the sub-models for the two non-user groups, namely, AUD outcome in group~C and CUD outcome in group~A, have only intercepts and no predictors. The same, however, is also true for the CUD outcome in group~C. The remaining three sub-models have six predictors each --- two (gender and delinquency) common to all, two (extraversion and race) common to groups A and B for AUD, and two (conscientiousness and neuroticism) common to AUD and CUD for group~B. The group~A-AUD sub-model has two unique predictors, namely, parental education and peer alcohol. Similarly, the group~B-CUD sub-model has openness and peer cannabis as unique predictors. On the other hand, the group~B-AUD sub-model does not have any unique predictor. Altogether the joint model has 10 predictors --- 8 for AUD, 6 for CUD, and 4 that are common. 

% Table 3

%%%%%%%%%%%%%%%%%%
\begin{table}
  \caption{Proposed Joint Model: Posterior means of regression coefficients in the model and odds ratios (ORs) and 95\% credible intervals (CIs) for ORs. The ORs are in the original scale of the predictor.}
   \label{t:ch4t6}
  \begin{center}
  \begin{tabular}{llrrlc} \hline
  & & \multicolumn{2}{c}{\textbf{Posterior mean}} && \textbf{95\% CI} \\ \cline{3-4} \cline{6-6}
 \textbf{Sub-model} & \textbf{Variable} & \textbf{Coefficient} &
   \textbf{OR} && \textbf{OR} \\   \hline
Group A-AUD &Intercept &	$-10.02$\\
   &Biological gender	(Male) &0.63 & 1.94 && (1.15, 3.13)\\
  &Parental education & 1.39 & 1.33 && (1.08, 1.63)\\
&Peer alcohol use& 1.11 & 1.13 && (1.03, 1.25)\\
&Extraversion scale&	2.24 & 1.12 && (1.03, 1.22)  \\
&Delinquency & 13.74 & 1.12 && (1.09, 1.15)\\
&Race (White) & 1.04 & 2.99 && (1.47, 5.66)\\[0.4em]
Group A-CUD &Intercept & $-17.21$\\[0.4em]
Group B-AUD & Intercept &	$-6.83$\\
   &Biological gender	(Male) & 0.76 & 2.16 && (1.66, 2.77)\\
  &Conscientiousness scale & $-2.19$ & 0.90 &&  (0.86, 0.94) \\
&Neuroticism scale& 2.20	& 1.12 && (1.07, 1.17)\\
&Extraversion scale&	2.20 & 1.12 && (1.08, 1.16)\\
&Delinquency & 6.34& 1.05 && (1.04, 1.06)	\\
& Race (White) & 1.36 & 3.95 && (2.86, 5.32) \\[0.4em]
Group B-CUD &Intercept &	$-7.44$\\
   & Biological gender (Male)	&0.52 & 1.70 && (1.26, 2.25)\\
  & Conscientiousness scale& $-1.69$& 0.92 && (0.87, 0.97)\\
& Neuroticism scale&	2.74 & 1.15 && (1.09, 1.21) \\
& Openness scale&	1.81 & 1.10 && (1.04, 1.15)\\
& Delinquency	& 6.66 & 1.06 && (1.05, 1.07)\\
& Peer cannabis use &	0.98 & 1.18 && (1.11, 1.25)\\[0.4em]
Group C-AUD &Intercept & $-14.98$ \\[0.4em]
Group C-CUD &Intercept &$-6.33$\\ \hline
    \end{tabular}
  \end{center}
\end{table}
%%%%%%%%%%%%%%%%%%%%%%%%%

Table~\ref{t:ch4t6} also provides posterior means of the regression coefficients in the proposed model. Recall that the continuous predictors in the model have been scaled to lie within $[0,1]$. However, the odds ratio (OR) associated with a coefficient is easier to interpret on the original scale of the predictor. This OR is given by $\text{exp}(\beta/M)$, where $M$ is the maximum value of the predictor used for scaling. The resulting ORs and their 95\% credible intervals are also presented in Table~\ref{t:ch4t6}. In group~A, higher likelihood of AUD is associated with male gender, white race, greater involvement in delinquent activities, higher extraversion, having parents with higher education levels, and greater peer alcohol use. The same holds true for the first four predictors in group~B-AUD. In addition, higher likelihood of AUD is associated with greater neuroticism and lower conscientiousness. In group~B, higher likelihood of CUD is associated with male gender, greater involvement in delinquent activities, greater openness and neuroticism, lower conscientiousness, and greater peer cannabis use. Lastly, the posterior mean and 95\% credible interval for $\sigma_u^2$ are 6.61 (5.81, 7.50).

\section{External Validation of the Proposed Joint Model}\label{se:validation} 

We validate the model on two independent test datasets: a subset of Add Health consisting of participants with missing survey weights (and hence not included in the training set) and CHDS \citep{Poulton2020}. For each participant in both datasets, we calculate the predicted probability of developing AUD and CUD using the appropriate sub-models of the proposed model. Then, the predictions are compared with the actual status of AUD and CUD  through the predictive accuracy measures AUC, E/O, and Brier score.

\subsection{Validation on Add Health Test Set}

This dataset has 785 participants. The lifetime prevalences of AUD and CUD in these data are 7.64\% and 5.22\%, respectively. The composition of this dataset by substance use groups (with group-specific lifetime AUD and/or CUD prevalence) 
is as follows: 216 participants in group A (AUD: 3.70\%), 531 participants in group B (AUD: 9.79\%, CUD: 7.72\%), and 38 participants in group C (CUD: 0\%). This composition differs somewhat from that of the training set, which was presented in Table~\ref{ta:new1}.

The AUCs for AUD and CUD outcomes are 0.748 and 0.710, respectively. The corresponding Brier scores are 0.066 and 0.045 and E/O values are 1.23 and 1.07. Supplement Tables S6 and S7   provide for user groups A and B, respectively, the E/O values for the five risk quintile groups and also for different levels of each predictor. For a continuous predictor, two levels --- above and below the median are considered. For user group A, the E/O values for AUD are within 20\% of the target value of 1 for the fourth risk quintile group, male gender, above median delinquency, and below median peer alcohol use. For the remaining predictor levels, the values do not deviate too far from 1 except for female gender, above median extraversion, and below median delinquency. For user group B, all E/O values for AUD are between 1.01 and 1.64, indicating that the model may be slightly overpredicting the number of AUD cases in these data. The E/O values for group~B-CUD range between 0.20 and 1.44, with about 2/3 of the values lying within 20\% of the target.

\subsection{Validation on CHDS Data}

CHDS collected data on the health, education, and life progress of a group of 1,265 children born in Christchurch, New Zealand in 1977. It was used by \citet{Ruberu2023} to validate a CUD risk prediction model. A detailed description of the data and their pre-processing can be found in their paper. Our validation set here consists of 795 participants with complete data on the predictors in the proposed model. The lifetime prevalences of AUD and CUD are 12.58\% and 13.21\%, respectively, which are higher than those for the Add Health training data. Supplement Table S5 summarizes how the risk factors were defined in this dataset and their comparison with the Add Health training set. The variable in CHDS that is closest to race is Maori ethnicity at birth, which is not informative with regard to the Add Health race categories. Therefore, we assign ``Other" race category to all CHDS participants.

The breakdown of CHDS data by substance use groups (with group-specific lifetime AUD and/or CUD prevalence) is as follows: 158 participants in group A (AUD: 1.90\%), 637 participants in group B (AUD: 15.23\%, CUD: 16.48\%), and 0 participants in group C, i.e., all participants are users of alcohol. As there is substantial prevalence disparity between the Add Health training data and the CHDS validation data and it is mainly due to the countries being different, we also explore a re-calibration of the model, as recommended 
%by \citep{VanCalster2019} and references therein.
in the literature \citep{Steyerberg2004, Janssen2008, Moons2012A, VanCalster2019}. 
Following \citet{Ruberu2023}, we apply logistic re-calibration to update the model's intercept to align with the validation data prevalence. This adjustment focuses solely on the intercept, leaving the coefficients of the predictors unchanged. It does not affect the AUC because the relative rankings of predicted probabilities among the individuals in the test dataset remain the same. See \citet{Ruberu2023} for additional details about the re-calibration method.

The AUCs for AUD and CUD outcomes are 0.65 and 0.75, respectively. The corresponding Brier scores before re-calibration are 0.115 and 0.146. They decrease to 0.105 and 0.126 after re-calibration. The corresponding E/O values before re-calibration are 0.28 and 0.30. After re-calibration, both E/O values become 1, showing substantial improvement in the model performance.  Supplement Table S8 presents for user group~B, the updated E/O values for the five risk quintile groups, two levels of biological gender and race, and above/below median predictor groups. All values are more or less close to 1. In user group A, there are only 3 cases and the overall E/O is 1.004.

Taken together, these results indicate good calibration and discrimination performance of the model on both Add Health test data and CHDS data. The external validation performance is in line with those of some of the widely used risk prediction models in practice \citep{spie:etal:1994,
cons:etal:1999, Spiegelman2001}.

\section{Simulation Study}\label{se:simulation}

The proposed statistical methodology considers a joint model for AUD and CUD with sub-models for user groups A, B, and C. A standard alternative is to consider separate univariate models for AUD and CUD with the AUD model trained on all alcohol users (i.e., groups A and B combined) and the CUD model trained on all cannabis users (i.e., groups B and C combined). Unlike the joint model, the univariate model for an SUD does not have group-specific sub-models. Hence the same model applies to all users of that specific substance irrespective of whether or not they use the other substance. We conduct a simulation study to compare the joint and univariate modeling approaches for predicting AUD status for alcohol users and CUD status for cannabis users. 

\subsection{Settings and Data Generation}

We simulate data by mimicking the steps that we carried out to fit the joint model on Add Health data. Specifically, we generate 21 potential predictors that we considered initially and assign their effects (regression coefficients) under four different settings. Settings 1 and 2 use non-zero coefficients for the predictors in the final joint model described in Section~\ref{se:results} (the two settings differ in sample sizes). As this model does not have any predictor for CUD in group C, in setting~3, we additionally allow four predictors in group C-CUD to have non-zero coefficients. Setting~4 is specifically constructed to favor univariate modeling wherein the chance of an SUD for a user of only that specific substance is the same as that for a user of both substances. This is achieved by setting the SUD predictors and their effects to be the same for the two types of users. Table~\ref{sim_setting} presents the regression coefficients for each setting for the predictors that have at least one non-zero coefficient. The coefficients for the remaining predictors are set to zero. Intercepts are chosen so that the proportion of SUD cases in each outcome-group combination is similar to that in the Add Health data. 

% Table 4
%%%%%%%%%%
\begin{sidewaystable}[htbp]
  \caption{Regression coefficients for each simulation setting. The coefficients for predictors not shown are set to zero.}
   \label{sim_setting}
  \begin{center}
  \begin{tabular}{llrrrcrrrcrrr} \hline
    & & \multicolumn{3}{c}{\textbf{Group A}}  && \multicolumn{3}{c}{\textbf{Group B}} && \multicolumn{3}{c}{\textbf{Group C}}\\  \cline{3-5} \cline{7-9} \cline{11-13}
Response & Predictor &  1--2  &  3 &  4 &
 & 1--2  &  3 &  4 && 1--2  &  3 &  4\\ \hline
AUD & Intercept & 10.0 & $-10.02$ & $-8.50$ && $-6.83$ & $-6.83$ & $-8.20$ && $-14.98$ & $-14.98$ & $-14.98$ \\
& Biological gender & 2.00 & 2.00 & 2.00 && 1.00 & 1.00 & 2.00  && 0 & 0 & 0\\
& Neuroticism scale & 0 & 0 & 0 && 3.00 & 3.00 & 0   && 0 & 0 &0\\
& Delinquency & 15.00 & 15.00 & 15.00 && 6.34 & 6.34 & 15.00 &&  0 & 0 & 0\\
& Conscientiousness scale & 0 & 0 & 0 && $-2.19$ & $-2.19$ &0   && 0 & 0 & 0\\
& Extraversion scale  & 2.50 & 2.50 & 2.30 && 2.50 & 2.50 & 2.30  && 0 &0&0 \\
& Race & 1.50 & 1.50 & 1.50  && 1.50 & 1.50 & 1.50 && 0 &0&0 \\ 
& Parental education & 2.00 & 2.00 & 0 && 0 & 0 & 0   && 0 & 0 & 0 \\
& Peer alcohol use & 1.50 & 1.50 & 0 && 0 & 0 & 0  && 0 &0&0 \\    \hline
CUD & Intercept & $-17.21$ & $-17.21$ & $-17.21$ && $-7.44$ & $-7.44$ & $-9.00$  && $-4.30$ &$-10.00$ & $-9.50$ \\
& Biological gender & 0 & 0 & 0 && 1.50 & 1.50 & 2.00  && 0 &1.00 &2.00 \\
& Neuroticism scale & 0 & 0 & 0 && 3.50 & 3.50 & 4.00  && 0 & 4.00 & 4.00 \\
& Delinquency & 0 & 0 & 0 && 6.66 & 6.66 & 10.00  && 0 & 15.00 & 10.00 \\ 
& Conscientiousness scale & 0 & 0 & 0  && $-1.69$ & $-1.69$ & 0  && 0 & 0 & 0 \\  
& Openness scale & 0 & 0 & 0 && 1.81 & 1.81 & 0  && 0 & 0 & 0 \\  
& Peer cannabis use & 0 & 0 & 0 && 1.50 & 1.50 & 2.00  && 0 & 2.00 & 2.00 \\    \hline
   \end{tabular}
  \end{center}
\end{sidewaystable}
%%%%%%%%%%%%

The row (design) vectors $\textbf{x}^1_{ij}$ and $\textbf{x}^2_{ij}$ consist of both continuous and categorical predictor values. The continuous predictors need to be on the 0-to-1 scale. %for the model coefficients to be applicable. Thus, to generate the continuous predictors, we do the following.
We generate them as follows. 
First, the values on the 0-to-1 scale (used in model training) are transformed into the logit scale and their mean vector and covariance matrix are obtained. Then, we simulate the predictor values from a multivariate normal distribution with these specific mean vector and covariance matrix. Finally, the simulated predictors on the logit scale are transformed to the 0-to-1 scale via the inverse logit transformation. The categorical predictors are simulated similarly from their joint distribution estimated from the Add Health training data. Using the simulated predictor values and the regression coefficients from Table~\ref{sim_setting}, we generate two continuous outcomes from the following bivariate normal (BVN) distribution: 
\[
\begin{pmatrix}
y^{1*}_{ij}\\
y^{2*}_{ij}
\end{pmatrix} \sim \text{BVN} \begin{pmatrix}
\begin{pmatrix}
{\textbf{x}^1_{ij}}^{\prime}\beta^1_j \\
{\textbf{x}^2_{ij}}^{\prime}\beta^2_j
\end{pmatrix}$, $\begin{pmatrix}
\sigma^2_{1j} & \sigma_{1j}\sigma_{2j}\rho_j \\
\sigma_{1j}\sigma_{2j}\rho & \sigma^2_{2j}
\end{pmatrix} 
\end{pmatrix}, 
\]
where $i = 1, \ldots, n_j$, $j = \text{A}, \text{B}, \text{C}$. We fix $\sigma^2_{1j}=\sigma^2_{2j}=5$ for all $j$, $\rho_B \in \lbrace 0.2, 0.8 \rbrace$, and $\rho_A = \rho_C = {0}$ (as only one outcome is possible in these groups). Finally, we dichotomized the simulated continuous outcomes as $y^k_{ij} = 0$ if $y^{k*}_{ij}\leq 0$, otherwise $y^k_{ij} = 1$ for $k = 1, 2$, to get the two binary outcomes.

We generate a training sample of the size mentioned in Supplement Table S9 under a specific setting. Then, we use it to build joint and univariate models with $t$ priors for AUD and CUD. Variable selection is performed using the threshold method with 0.10 cutoff. We evaluate the predictive accuracy of the models on two test datasets of varying size as indicated in Supplement Table S9 and generated in the same manner as the training data for that setting. As settings~1 and 2 differ only in the size of the training dataset, the test datasets for these settings are the same. 

We repeat the above process of generating a sample, building models, and using them to do prediction on the test datasets 500 times under each setting. The average values of predictive accuracy metrics AUC, Brier score, and E/O value are computed. 
%We also compute the proportion of times a model correctly identifies all associated predictors, i.e., those with non-zero regression coefficients (as in Supplement Table~\ref{sim_setting}). For a joint model, correct identification refers to including in its group-specific sub-models, the corresponding associated predictors all together. For a univariate model, correct identification refers to including all predictors associated with that SUD irrespective of whether it is relevant for user of that substance only or both substances.
%For a univariate model, correct identification refers to including the union set of group-specific predictors with non-zero coefficients.

\subsection{Results}

\begin{figure}[htbp]
\begin{center}
\includegraphics[scale=1] {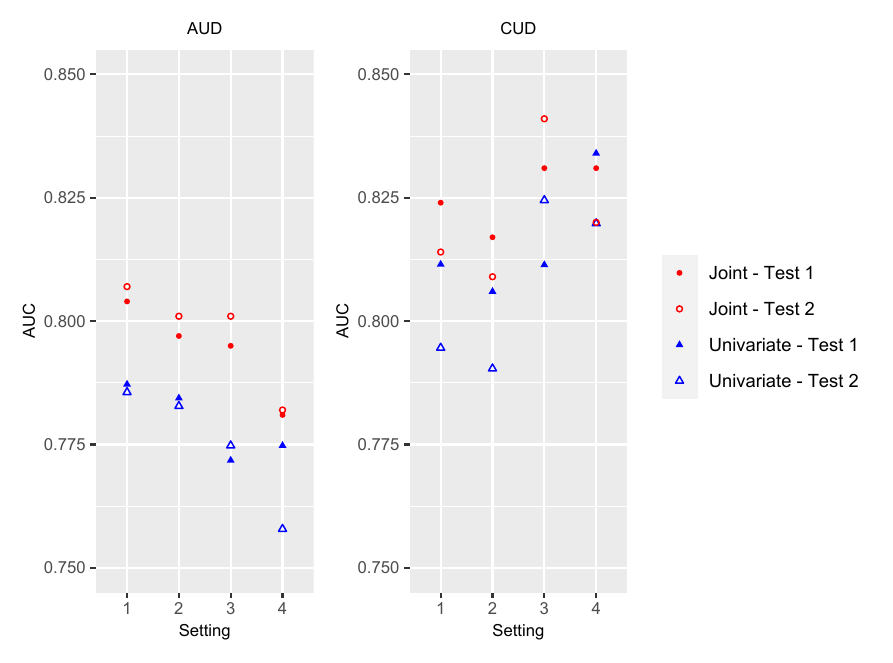}
\caption{Simulation Results: Comparison of AUC values for joint and univariate models.}
\label{fig:AUC_values}
\end{center}
\end{figure}

\begin{figure}[htbp]
\begin{center}
\includegraphics [scale=1] {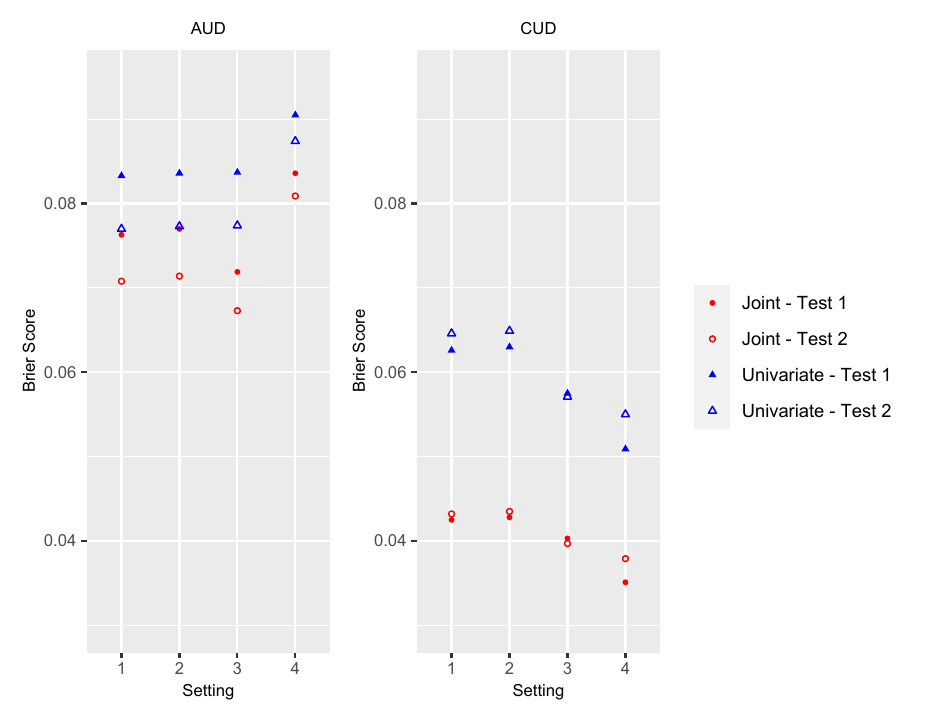}
\caption{Simulation Results: Comparison of Brier scores for joint and univariate models.}
\label{fig:brier_score}
\end{center}
\end{figure}

Figures~\ref{fig:AUC_values} and~\ref{fig:brier_score} display the average AUC and Brier score values for $\rho_B=0.8$. The AUC values of the joint model are greater than their univariate counterparts in all cases except one:  CUD under setting~4 with test set~1. But even in this case, the values are practically the same (0.834 for univariate models vs 0.831 for joint model). The corresponding values for test set~2 are identical at 0.820. The Brier scores for the joint model are smaller than their univariate counterparts for all settings and both test sets. The E/O values for the joint and univariate models tend to be similar and do not exhibit any clear pattern (results not presented). Qualitatively similar results are observed for $\rho_B=0.2$. Overall, these findings indicate superiority of joint modeling over univariate modeling in terms of predictive accuracy. Recall that setting~4 is specifically designed to favor univariate modeling. So it is notable that even under this setting, joint modeling either outperforms univariate modeling or matches its performance.

% Figure 1

%%%%%%%%%%%%%%%%%%%%%%%%
%%%%%%%%%%%%%%%%%%%%%%%%

% Figure 2

%%%%%%%%%%%%%%%%%%%%%%%%
%%%%%%%%%%%%%%%%%%%%%%%%

%Figure~\ref{fig:imp_pred} displays the percentage of times joint and univariate models identify all associated predictors in each setting. The percentages decrease from setting~1 to~2 as the sample size of the training set decreases. For both outcomes and in all settings, the percentages for the group~B sub-model of the joint model are at least as large as those for the univariate model. AUD univariate model has 100\% identification rate in setting~4, which far exceeds the rates in settings~1-3 because setting~4 is specifically designed to favor univariate modeling. The percentages for the group~A sub-model are smaller compared to those for the group~B sub-model (in all settings) and the univariate model in setting~4. This is partly explained by the fact that group~A has smaller sample size compared to group~B as well as the univariate model. For CUD, the group~C sub-model either does not have any associated predictors (settings~1 and~2) or does not identify all associated predictors together (settings 3 and 4), most likely due to the smallest sample size of this group. In addition, under settings~3 and~4, the percentages for the group~B sub-model and the univariate model are similar. This may be partly due to the fact that the CUD predictors in groups~B and C overlap and the univariate model is trained on combined data from groups~B and~C. Altogether, these findings indicate that joint modeling may be considered better than univariate modeling for identifying all associated predictors, especially if they vary with user groups. 
\section{Discussion}\label{se:discussion}

We proposed a novel Bayesian learning methodology to jointly model two dependent outcomes across different groups of participants. The model is highly flexible as it: (i) can accommodate both common and unique predictors across different groups and outcomes; (ii) allows the effects of the shared predictors to vary across different group-outcome combinations; (iii) allows one or more group-outcome combinations to have no predictors if that is necessitated by the task at hand; and (iv) is applicable to anyone who is at risk of {\em at least} one of the outcomes, thereby making the model applicable to a larger target population. The methodology is applied to build a model for predicting the probability of developing CUD and AUD in adulthood using predictors measured in adolescence/youth. We train the model on a large, nationally representative longitudinal dataset. Its prediction accuracy is satisfactory in both internal and external validation. The validation on CHDS data from an entirely different country is especially remarkable, demonstrating the robustness and portability of the model. Our simulation study confirms that the joint modeling of dependent outcomes is a better choice than separate univariate modeling on the basis of prediction accuracy.
%and the ability to identify all associated predictors, especially if the predictors and their effects vary across user groups.

As described in Supplement Section S2, 
in our prediction accuracy calculation for the joint model, we excluded AUD predictions for non-users of alcohol and CUD predictions for non-users of cannabis. This is to reflect the fact that in real clinical settings, prediction of risk of developing an SUD for a non-user of that substance is not of an interest. If we had included those predictions, the AUC values would have been higher because our model gives zero probability for non-users, which matches perfectly with their observed SUD value of 0. 

An important consideration for adopting a risk prediction model in practice is model parsimony and simplicity. This is especially true for a sensitive target population such as the adolescent substance users. To get a parsimonious model, we excluded ACE because its credible interval %(for group B-AUD) 
included zero. This exclusion does not affect the prediction accuracy because ACE's effect is captured to an extent by other predictors such as neurotisicm and race. For model simplicity, we explored  dichotomizing race (white vs non-white) following evidence in the literature that white race is associated with higher AUD risk. \nocite{Chartier2010} The dichotomous version leads to a simpler model with the same prediction accuracy. Model simplicity also prompted us to consider, for each longitudinal predictor, using average across different waves rather than the random effects. This exercise not only resulted in a model that is easier to use and interpret but also has higher predictive accuracy.

The proposed joint model has just 10 unique predictors to model the two SUD outcomes for three different user groups.
Some of these predictors are common across the outcomes and the groups such as gender, delinquency, neuroticism, and conscientiousness. Others such as extraversion and race are there only for AUD. There are also group-outcome specific predictors, e.g., peer alcohol use and parental education for group~A-AUD; and peer cannabis use and openness for group~B-CUD.

The direction of association between the predictors and the SUD outcomes in the proposed model are consistent with the literature. Specifically, it is known that males are more likely to develop SUD than females \citep{Hayatbakhsh2009,Meier2016}. %,Agabio2017}
%Several studies have reported that 
Involvement in delinquent activities increases the likelihood of developing SUDs \citep{Koh2017, Mason2010, Jing2020}. 
Moreover, the personality variables, including neuroticism and conscientiousness, have been previously identified as important in predicting substance use and dependence \citep{Cooper2000, Zoboroski2021,
Rajapaksha2022}. Previous studies have also found peer alcohol use, white race, and extraversion as important for alcohol-related outcomes \citep{Fairbairn2015, Goldstick2019, Chena2004} %NIAAA2022
A seemingly counter-intuitive finding is that adolescents or young adults in group A with more educated parents are more likely to develop AUD. Nonetheless, this is consistent with the finding of positive association between high alcohol use and high socioeconomic status of adolescents as measured through parental education  \citep{Humensky2010, Patrick2012}.

%Nonetheless, it is consistent with a study that reported there is evidence of positive association of high socioeconomic status of adolescents as measured through parental education with high alcohol use \citep{Humensky2010, Patrick2012}.

%No predictor was identified as important for the CUD outcome in group~C (cannabis-only users). 

Our model did not identify any predictor as important for the CUD outcome in group~C (cannabis-only users). This may be due to the small sample size of the group and the low CUD prevalence in it, which are both supported by the literature. Specifically, cannabis users tend to use other substances as well, especially alcohol \citep{Yurasek2017, %Hyggen2015, %MIDANIK200772
Roche2019}. Moreover, cannabis users who use other substances are more likely to get CUD compared to those who use only cannabis \citep{Roche2019}. Conversely, users of substances other than cannabis, e.g., alcohol, are also likely to be users of cannabis  \citep{Weinberger2016}. Altogether these results also explain why group~B has a larger sample size and higher prevalence of AUD and CUD than groups~A and~C. 

There are some limitations of our study. First, the Add Health data are relatively old and do not capture the recent changes, especially the increasing prevalence of cannabis use due to easier access \citep{Knapp2019%,Spindle2019
}. Nonetheless, the set of personal risk factors that makes a cannabis user susceptible to CUD are not likely to change substantially with the easier availability of cannabis. Next, Add Health did not collect data on neurocognitive and neuroimaging variables which may also be important for predicting SUD. Moreover, the personality related variables in the final model were measured only in wave~IV. However, there is evidence in the literature that personality traits remain relatively stable over time \citep{Caspi2005,Damian2019}.

These limitations are outweighed by the various strengths of the study. Our Bayesian learning methodology is novel from both the methodological and application points of view. Moreover, the final model has good discrimination and calibration performance on two external validation datasets, including one from a different country. The flexible modeling framework proposed in this study can also be used to develop risk prediction models for other SUD outcomes jointly, e.g., CUD and tobacco use disorder, or other diseases, e.g., joint modeling of two related cancers.

In summary, our proposed model can help in identifying adolescent or young adult substance users who are at a high risk of developing SUD in adulthood. These users can be provided with appropriate intervention
or prevention measures to help them steer away from the path
towards SUD. Successful implementation of the model in actual practice though will require careful considerations in identifying appropriate clinical settings and reaching out to the target population. %as discussed in \citet{Ruberu2023}.

\section*{Acknowledgements}
This work was partially funded by the University of Texas at Dallas SPIRe seed grant. CHDS is funded by the Health Research Council of New Zealand, Programme Grant 16/600.

\bibliography{thesis_new_v2}

\clearpage
\renewcommand{\baselinestretch}{1}

%%%%%%%%%%%%%%%%%%%%%%%%%%
% Tables
%%%%%%%%%%%%%%%%%%%%%%%%%%

\clearpage

%%%%%%%%%%%%%%%%
% Figures
%%%%%%%%%%%%%%%



\section*{Supplementary material}

# Supplementary Materials for

# A Bayesian Learning Model for Joint Risk Prediction of Alcohol and Cannabis Use Disorders

Rajapaksha Mudalige Dhanushka S. Rajapaksha, Tingfang Wang, Thanthirige Lakshika M. Ruberu, Joseph M. Boden, Pankaj K. Choudhary, and Swati Biswas

# S1 Add Health Training Data: Preprocessing and Sample Characteristics

## S1.1 Data Proprocessing

Add Health conducted its first survey (wave I) during 1994–95 school year when the participants were in grades 7–12 (Harris et al., 2009). It selected participants using a multistage stratified cluster sampling design to ensure that they were representative of the then US adolescent population in terms of urbanicity, region, school size and type, and ethnicity. Follow-up surveys were conducted in 1996 (wave II), 2001–02 (wave III), 2008 (wave IV), and 2016–18 (wave V). No substance use related information was collected in the last wave. So the relevant data for our purpose come only from the waves I–IV during which the participants became adults aged 24–32. The outcome variables — binary measures of lifetime AUD and CUD diagnosed using the 4th edition of the Diagnostic and Statistical Manual of Mental Disorders (Marel et al., 2019) — were measured only in wave IV.

A participant was included in our training data if they satisfied the following criteria: began alcohol or cannabis use during the first three waves, took part in wave IV, had alcohol/cannabis use status across the different waves consistent with the age of first alcohol/cannabis use, and had survey weights available. Additionally, for those who developed AUD or CUD, the corresponding age of onset was available. This age was defined as the age of AUD onset in group A, age of CUD onset in group C, and the minimum of the

two ages in group B. The survey data were processed along the lines of Rajapaksha et al. (2022). Initially, about 50 potential predictors were constructed from the survey questionnaires based on a review of the literature. Some predictors were cross-sectional, measured in a single wave, and others were longitudinal, measured in multiple waves. Each longitudinal predictor was summarized in two ways (Chen et al., 2015; Dandis et al., 2020): (a) the estimated random intercept obtained by fitting a linear mixed effects model (LMM) relating the predictor to the participant age at different waves; and (b) the average value over the different waves. All quantitative predictors were transformed to lie within [0,1] to make their scales comparable. For those who developed AUD or CUD, only the portion of their data that were recorded chronologically before their age of onset was considered. This ensured that all predictors were measured before the outcome. Since some predictors had a rather large proportion of data missing, the variable filtering process implemented in Rajapaksha et al. (2022) to maximize the number of participants with complete data without losing any potentially important predictors was applied twice separately for AUD and CUD. This process resulted in 15 predictors for AUD and 17 predictors for CUD — altogether a total of 21 unique predictors — with complete data on $n = 12,503$ participants. This dataset forms our training set.

## S1.2   Sample Characteristics

Table S1: List of predictors that passed the initial variable filtering.

| Predictor | Description |
| --- | --- |
| Biological gender | Female, male |
| Race | White, Black, Native American, Asian, other |
| Neuroticism scale | Measure of general tendency to experience negative feelings. A higher value implies greater tendency. |
| Conscientiousness scale | Measure of forward planning, organization, and ability to carry out tasks. A higher value implies greater conscientiousness. |
| Agreeableness scale | Measure of compassion, eagerness to cooperate, and tendency to avoid conflict. A higher value implies more agreeableness. |
| Openness scale | Measure of openness to new experiences and imaginativeness. A higher value implies more openness. |
| Extraversion scale | Measure the extent to which an individual is outgoing, social, and seeks stimulation from the external environment. A higher value implies more extraversion. |
| Physical health | Measures the frequency of minor physical health problems. A higher value implies worse health condition. |
| Anxiety | Number of times experienced anxiety symptoms |
| Depression | Number of times experienced depression symptoms |
| Delinquency | Number of times involved in delinquent activities |
| Self esteem scale | Measure of individual's subjective overall evaluation of their own worth. A higher value implies lower self esteem. |
| ACE | Number of adverse childhood experiences that occurred before age 18 |
| High school behavior | Measure of school-related problems not experienced by respondent. A higher value implies fewer problems. |
| Violence victimization | Number of times experienced violent incidents |
| Supportive environment | Measure of perception of being supported by parents, teachers, friends, etc. A higher value implies greater support. |
| Peer alcohol use | Number of best friends (out of 3) who drink alcohol at least once a month |
| Peer cannabis use | Number of best friends (out of 3) who use cannabis at least once a month |
| Peer smoking | Number of best friends (out of 3) who smoke at least 1 cigarette a day |
| Parental education | Measure of highest education level achieved by any of the parents. A higher value implies more educated parents. |
| TV hours | Number of hours watched television per week |

Table S2: Summary of cross-sectional predictors selected for AUD and CUD: percentage for categorical predictors; mean (standard deviation) for continuous predictors; and p-values for a chi-square test of association for a categorical predictor and a univariate logistic regression for a continuous predictor.

|  | | AUD | | | | CUD | | |
| --- | --- | --- | --- | --- | --- | --- | --- | --- |
|  | Overall | Case | Control | P-value | | Case | Control | P-value |
| Predictor | (n=12503) | (n=1184) | (n=11319) | | | (n=636) | (n=11867) | |
| Biological gender | | | | < 0.001 | | | | < 0.001 |
| Male | 47.77 | 60.39 | 46.45 | | | 61.16 | 47.05 | |
| Female | 52.23 | 39.61 | 53.55 | | | 38.84 | 52.95 | |
| Race | | | | < 0.001 | | | | 0.254 |
| White | 68.99 | 82.43 | 67.59 | | | 72.17 | 68.82 | |
| Black | 20.77 | 9.80 | 21.92 | | | 19.97 | 20.81 | |
| Native American | 1.83 | 1.60 | 1.86 | | | 2.04 | 1.82 | |
| Asian | 6.61 | 4.90 | 6.79 | | | 4.25 | 6.74 | |
| Other | 1.79 | 1.27 | 1.85 | | | 1.57 | 1.80 | |
| ACE scale | 0.20 (0.18) | 0.23 (0.18) | 0.20 (0.18) | < 0.001 | | 0.26 (0.19) | 0.20 (0.17) | < 0.001 |
| TV hours | 0.16 (0.15) | 0.15 (0.14) | 0.16 (0.15) | 0.025 | | 0.17 (0.15) | 0.16 (0.15) | 0.185 |
| Neuroticism scale | 0.52 (0.14) | 0.55 (0.15) | 0.52 (0.14) | < 0.001 | | 0.56 (0.14) | 0.52 (0.14) | < 0.001 |
| Conscientiousness scale | 0.73 (0.13) | 0.70 (0.15) | 0.73 (0.13) | < 0.001 | | 0.70 (0.14) | 0.73 (0.13) | < 0.001 |
| Openness scale | 0.73 (0.12) | 0.75 (0.13) | 0.73 (0.12) | < 0.001 | | 0.76 (0.13) | 0.73 (0.12) | < 0.001 |
| Agreeableness scale | 0.76 (0.12) | | | | | 0.77 (0.12) | 0.76 (0.12) | 0.751 |
| Extraversion scale | 0.67 (0.15) | 0.70 (0.15) | 0.66 (0.15) | < 0.001 | | | | |
| High school trouble | 0.27 (0.18) | 0.31 (0.19) | 0.26 (0.18) | < 0.001 | | | | |
| Parental education | 0.64 (0.24) | 0.69 (0.23) | 0.64 (0.24) | < 0.001 | | | | |

Table S3: Summary of longitudinal predictors selected for AUD outcome: mean (standard deviation) across waves in which they were measured. For each predictor, the p-value based on a univariate logistic regression model with AUD status as response and the estimated random intercept associated with the predictor as covariate is $< 0.001$.

| Predictor | Wave 1 | | | Wave 2 | | | Wave 3 | | |
|---|---|---|---|---|---|---|---|---|---|
| | Overall | Case | Control | Overall | Case | Control | Overall | Case | Control |
| Delinquency | 0.10 | 0.14 | 0.10 | 0.07 | 0.11 | 0.07 | 0.02 | 0.04 | 0.02 |
| | (0.12) | (0.14) | (0.11) | (0.10) | (0.12) | (0.09) | (0.05) | (0.07) | (0.05) |
| Violence | 0.04 | 0.05 | 0.04 | 0.02 | 0.02 | 0.02 | 0.01 | 0.02 | 0.01 |
| victimization | (0.07) | (0.08) | (0.07) | (0.05) | (0.06) | (0.05) | (0.05) | (0.05) | (0.05) |
| Peer alcohol | 0.39 | 0.45 | 0.39 | 0.40 | 0.48 | 0.39 | 0.60 | 0.76 | 0.60 |
| use | (0.39) | (0.40) | (0.39) | (0.39) | (0.39) | (0.39) | (0.40) | (0.35) | (0.40) |
| Self esteem | 0.39 | 0.40 | 0.39 | 0.37 | 0.38 | 0.37 | 0.36 | 0.38 | 0.36 |
| scale | (0.13) | (0.13) | (0.13) | (0.13) | (0.12) | (0.13) | (0.11) | (0.11) | (0.11) |
| Supportive | 0.81 | 0.80 | 0.81 | 0.82 | 0.80 | 0.82 | | | |
| environment | (0.13) | (0.12) | (0.13) | (0.13) | (0.12) | (0.13) | | | |
| Physical | 0.20 | 0.22 | 0.20 | 0.20 | 0.22 | 0.20 | | | |
| Health | (0.10) | (0.10) | (0.10) | (0.09) | (0.09) | (0.09) | | | |

S5

Table S4: Summary of longitudinal predictors selected for CUD outcome: mean (standard deviation) across different waves in which they were measured. For each predictor, the p-value based on a univariate logistic regression model with CUD status as response and the estimated random intercept associated with the predictor as covariate is $< 0.001$.

| Predictor | Wave 1 | | | Wave 2 | | | Wave 3 | | |
|---|---|---|---|---|---|---|---|---|---|
| | Overall | Case | Control | Overall | Case | Control | Overall | Case | Control |
| Anxiety | 0.14 | 0.16 | 0.14 | 0.14 | 0.16 | 0.14 | | | |
| | (0.10) | (0.11) | (0.11) | (0.10) | (0.10) | (0.10) | | | |
| Depression | 0.27 | 0.28 | 0.27 | 0.27 | 0.28 | 0.27 | 0.29 | 0.32 | 0.29 |
| | (0.09) | (0.10) | (0.09) | (0.09) | (0.09) | (0.09) | (0.11) | (0.11) | (0.11) |
| Delinquency | 0.10 | 0.15 | 0.10 | 0.07 | 0.12 | 0.07 | 0.02 | 0.05 | 0.02 |
| | (0.12) | (0.14) | (0.11) | (0.10) | (0.13) | (0.10) | (0.05) | (0.07) | (0.05) |
| Violence victimization | 0.04 | 0.06 | 0.04 | 0.02 | 0.03 | 0.02 | 0.01 | 0.02 | 0.01 |
| | (0.07) | (0.08) | (0.07) | (0.05) | (0.07) | (0.05) | (0.05) | (0.06) | (0.05) |
| Peer alcohol use | 0.39 | 0.48 | 0.39 | 0.40 | 0.47 | 0.40 | 0.60 | 0.72 | 0.60 |
| | (0.39) | (0.41) | (0.39) | (0.39) | (0.39) | (0.39) | (0.40) | (0.37) | (0.40) |
| Peer cannabis use | 0.21 | 0.34 | 0.21 | 0.25 | 0.40 | 0.25 | | | |
| | (0.33) | (0.39) | (0.33) | (0.35) | (0.40) | (0.34) | | | |
| Peer smoking | 0.28 | 0.35 | 0.27 | 0.31 | 0.42 | 0.31 | | | |
| | (0.36) | (0.38) | (0.35) | (0.37) | (0.41) | (0.37) | | | |

# S2 Joint Bayesian Modeling: Variable Selection and Model Comparison

Both the variable selection methods — credible interval and thresholding (Van Erp et al., 2019; Li and Lin, 2010) — begin by fitting the full model with all predictors. Thereafter, the first one selects a predictor if the credible interval of its coefficient at a given level excludes zero. Whereas, the second one selects a predictor if the posterior probability of its coefficient falling within one posterior SD of zero is less than a given threshold. More variables are selected as the level in the first method increases or the threshold in the second method decreases. Here level and threshold serve as tuning parameters lying in $(0, 1)$ range. We vary them over a grid of values, e.g., in increments of 0.10, resulting in competing models, which are then compared to identify the best model.

We compare models on the basis of accuracy in predicting AUD and CUD for future individuals, estimated via 5-fold cross-validation (CV). The predicted probabilities obtained in the CV step are used to evaluate two aspects of model performance: discrimination and calibration. Discrimination is assessed by area under the receiver operating characteristic curve (AUC). Calibration is evaluated by comparing the expected number of cases (E) with their observed number (O) through the ratio E/O and the Brier score — mean squared difference between predicted probabilities and observed outcomes. Higher AUC, smaller Brier score, or E/O closer to one indicate better model performance. The three measures are computed separately for AUD and CUD. Since, in practice, risk prediction for a specific SUD will be carried out only for the users of that substance, we do not consider AUD predictions for non-users of alcohol and CUD predictions for non-users of cannabis in the predictive accuracy calculations.

# S3 CHDS Data: Risk Factors

Table S5: Comparison of predictors in Add Health training data and CHDS test data.

| Predictor | Add Health | CHDS |
|---|---|---|
| Biological gender | Female, male | Available |
| Race | White, Black, Native American, Asian, Other | New Zealand Maori ethnicity at birth (yes, no) |
| Conscientiousness, neuroticism, openness, and extraversion scale | Personality questionnaire where each scale is measured by 4 questions. | Neuroticism and extraversion were assessed at age 14 using a concise version of the Eysenck Personality Inventory (Eysenck and Eysenck, 1964). Openness was evaluated at age 16, utilizing items aligned with the Tri-dimensional Personality Questionnaire (Cloninger, 1987). Conscientiousness was measured at age 40 with a shortened personality assessment tool (Sibley, 2012). |
| Delinquency | Number of times involved in delinquent activities. A higher value implies greater involvement. | From ages 7-9, parents and teachers provided reports, combining elements from Rutter (Rutter et al., 1970) and Conners (Conners, 1969, 1970) questionnaires. Starting from age 10 to 16, the score included input from parents, teachers, and the participants themselves, employing a questionnaire based on items from the Diagnostic Interview Schedule for Children (Costello et al., 1985). |
| Peer alcohol use | Number of best friends (out of 3) who drink alcohol at least once a month | At ages 15 and 16 ("How many of your friends drink alcohol regularly"); at ages 18, 21, and 25 ("How many of your male/female friends drink alcohol regularly"). |
| Peer cannabis use | Number of best friends (out of 3) who use cannabis at least once a month | At ages 15 and 16 ("Whether your best friend or other close friends used cannabis in the last year"); at ages 18, 21, and 25 ("How many of your male/female friends use marijuana or other drugs"). |
| Parental education | Measure of highest education level achieved by any parent. A higher value implies more educated parents. | Based on parental education level at birth of the participant (no qualifications, secondary qualifications, tertiary qualifications). |

# S4 External Validation: Detailed Results

Table S6: Expected (E) and observed (O) number of AUD cases and E/O values in group A of Add Health test set.

|  |  | AUD | | |
|---|---|---|---|---|
| Group | n | E | O | E/O |
| Risk Quintile 1 | 44 | 0.47 | 0 | |
| Risk Quintile 2 | 43 | 0.79 | 1 | 0.79 |
| Risk Quintile 3 | 43 | 1.17 | 2 | 0.58 |
| Risk Quintile 4 | 43 | 1.85 | 2 | 0.93 |
| Risk Quintile 5 | 43 | 5.24 | 3 | 1.75 |
| Biological gender | | | | |
|   Male | 84 | 5.01 | 6 | 0.84 |
|   Female | 132 | 4.51 | 2 | 2.26 |
| Extraversion scale | | | | |
|   Below median | 112 | 3.12 | 6 | 0.52 |
|   Above median | 104 | 6.40 | 2 | 3.20 |
| Delinquency | | | | |
|   Below median | 108 | 2.68 | 1 | 2.68 |
|   Above median | 108 | 6.84 | 7 | 0.98 |
| Race | | | | |
|   White | 117 | 6.88 | 4 | 1.72 |
|   Non-white | 99 | 2.65 | 4 | 0.66 |
| Peer alcohol use | | | | |
|   Below median | 140 | 4.03 | 4 | 1.01 |
|   Above median | 76 | 5.49 | 4 | 1.37 |
| Parental education | | | | |
|   Below median | 133 | 4.67 | 4 | 1.17 |
|   Above median | 83 | 4.86 | 4 | 1.21 |

Table S7: Expected (E) and observed (O) number of AUD cases and E/O values in group B of Add Health test set.

|  |  | AUD | | | CUD | | |
|---|---|---|---|---|---|---|---|
| **Group** | n | E | O | E/O | E | O | E/O |
| Risk Quintile 1 | 107 | 4.64 | 4 | 1.16 | 3.28 | 5 | 0.66 |
| Risk Quintile 2 | 106 | 7.84 | 5 | 1.57 | 4.71 | 5 | 0.94 |
| Risk Quintile 3 | 106 | 10.80 | 8 | 1.35 | 6.34 | 6 | 1.06 |
| Risk Quintile 4 | 106 | 14.79 | 9 | 1.64 | 9.23 | 7 | 1.32 |
| Risk Quintile 5 | 106 | 26.32 | 26 | 1.01 | 19.58 | 18 | 1.09 |
| Biological gender | | | | | | | |
|   Male | 274 | 40.12 | 32 | 1.25 | 26.98 | 29 | 0.93 |
|   Female | 257 | 24.27 | 20 | 1.21 | 16.16 | 12 | 1.35 |
| Conscientiousness scale | | | | | | | |
|   Below median | 309 | 42.16 | 38 | 1.11 | 28.02 | 28 | 1.00 |
|   Above median | 222 | 22.23 | 14 | 1.59 | 15.12 | 13 | 1.16 |
| Neuroticism scale | | | | | | | |
|   Below median | 268 | 28.99 | 27 | 1.07 | 17.74 | 18 | 0.99 |
|   Above median | 263 | 35.40 | 25 | 1.42 | 25.40 | 23 | 1.10 |
| Delinquency | | | | | | | |
|   Below median | 267 | 22.93 | 15 | 1.53 | 12.84 | 13 | 0.99 |
|   Above median | 264 | 41.46 | 37 | 1.12 | 30.30 | 28 | 1.08 |
| Openness scale | | | | | | | |
|   Below median | 337 | 39.18 | 28 | 1.40 | 24.66 | 19 | 1.30 |
|   Above median | 194 | 25.21 | 24 | 1.05 | 18.48 | 22 | 0.84 |
| Peer cannabis use | | | | | | | |
|   Below median | 301 | 30.65 | 25 | 1.23 | 15.96 | 22 | 0.73 |
|   Above median | 230 | 33.74 | 27 | 1.25 | 27.18 | 19 | 1.43 |
| Race | | | | | | | |
|   White | 345 | 50.29 | 41 | 1.23 | 27.28 | 30 | 0.91 |
|   Non-white | 186 | 14.09 | 11 | 1.28 | 15.86 | 11 | 1.44 |
| Extraversion scale | | | | | | | |
|   Below median | 330 | 36.95 | 26 | 1.42 | 26.55 | 28 | 0.95 |
|   Above median | 201 | 27.44 | 26 | 1.06 | 16.59 | 13 | 1.28 |

Table S8: Expected (E) and observed (O) number of AUD cases in group B of CHDS data.

| Risk quintile groups | n | CUD E | CUD O | CUD E/O | AUD E | AUD O | AUD E/O |
|---|---|---|---|---|---|---|---|
| Risk Quintile 1 | 127 | 9.23 | 4 | 2.31 | 9.66 | 14 | 0.69 |
| Risk Quintile 2 | 127 | 13.97 | 7 | 2.0 | 13.95 | 13 | 1.07 |
| Risk Quintile 3 | 127 | 18.7 | 20 | 0.94 | 17.57 | 15 | 1.17 |
| Risk Quintile 4 | 127 | 24.44 | 31 | 0.79 | 22.58 | 22 | 1.03 |
| Risk Quintile 5 | 128 | 38.63 | 43 | 0.90 | 33.19 | 33 | 1.01 |
| Biological gender | | | | | | | |
|   Male | 319 | 59.57 | 74 | 0.81 | 58.47 | 53 | 1.10 |
|   Female | 318 | 45.43 | 31 | 1.47 | 38.53 | 44 | 0.88 |
| Conscientiousness scale | | | | | | | |
|   Below median | 410 | 76.12 | 73 | 1.04 | 70.79 | 70 | 1.01 |
|   Above median | 226 | 28.74 | 32 | 0.90 | 26.06 | 27 | 0.97 |
| Neuroticism scale | | | | | | | |
|   Below median | 241 | 34.43 | 35 | 0.98 | 33.50 | 32 | 1.05 |
|   Above median | 300 | 58.72 | 60 | 0.98 | 50.83 | 51 | 1.00 |
| Delinquency | | | | | | | |
|   Below median | 318 | 36.42 | 31 | 0.17 | 36.93 | 39 | 0.95 |
|   Above median | 318 | 68.53 | 74 | 0.93 | 60 | 58 | 1.03 |
| Openness scale | | | | | | | |
|   Below median | 336 | 44.75 | 37 | 1.21 | 47.35 | 41 | 1.15 |
|   Above median | 299 | 60.06 | 67 | 0.90 | 49.37 | 56 | 0.88 |
| Peer cannabis use | | | | | | | |
|   Below median | 326 | 45.07 | 35 | 1.29 | 46.10 | 40 | 1.15 |
|   Above median | 286 | 57.30 | 67 | 0.86 | 47.69 | 51 | 0.94 |
| Race | | | | | | | |
|   Other | 637 | 105 | 105 | 1.00 | 97 | 97 | 1.00 |
| Extraversion scale | | | | | | | |
|   Below median | 329 | 53.7 | 48 | 1.12 | 45.28 | 45 | 1.01 |
|   Above median | 306 | 50.83 | 57 | 0.89 | 51.45 | 52 | 0.99 |

# S5 Simulation Study: Settings

Table S9: Sizes of training and test datasets.

| Setting | Size | Training set | Test set 1 | Test set 2 |
|---|---|---|---|---|
| 1 | $n_A$ | 2500 | 5000 | 1250 |
|   | $n_B$ | 5000 | 10000 | 2500 |
|   | $n_C$ | 500 | 1000 | 250 |
| 2 | $n_A$ | 1250 | 5000 | 1250 |
|   | $n_B$ | 2500 | 10000 | 2500 |
|   | $n_C$ | 250 | 1000 | 250 |
| 3 | $n_A$ | 2500 | 5000 | 1250 |
|   | $n_B$ | 5000 | 10000 | 2500 |
|   | $n_C$ | 1000 | 2000 | 500 |
| 4 | $n_A$ | 2500 | 5000 | 1250 |
|   | $n_B$ | 5000 | 10000 | 2500 |
|   | $n_C$ | 500 | 1000 | 500 |



\end{document}